# Exact Solution to Terzaghi's Consolidation Equation

Romolo Di Francesco
GEO&GEO Instruments® - research & development
Teramo (TE), Italy
E-mail:    romolo.difrancesco@vodafone.it

**Abstract.**
The application of the consolidation equation is based on Taylor's approximate solution alone. The existence of the exact solution emerged from the analysis of the logical structure of d'Alambert's, Fourier' and Laplace's differential equations. This led to a nonlinear equation - based on the properties of elastic waves and elastic functions - which is able to simulate excess pore pressure transmission in the soil. The research is completed with the application of the solution obtained, thereby discovering that consolidation decay times may be calculated both through the construction of dissipation curves and through he analytical research of the time value satisfying the condition $\Delta u(z, t_{100}) = 0$. Finally, decay times match the approximate solution eliminating in fact the introduction of Taylor's additional parameters.

## 1  Introduction

Although the one-dimensional consolidation theory by Terzaghi & Fröhlich (1936) constitutes the basis of Soil Mechanics, it has never been fully applied, as Taylor's approximate solution (1948) is the only convenient solution. The existence of the exact solution to Terzaghi's equation emerged from researches regarding the description of geotechnical phenomena through direct or inverse hyperbolic laws (Di Francesco R., 2008b); the shift of these elements in Terzaghi's equation then led to a critical review of further studies (Di Francesco R., 2008a) on the analysis of d'Alambert's differential equation, on elastic wave propagation, and Fourier's differential equation, on heat transfer, considering the analogy of Fourier's differential equation with Terzaghi's equation and its reduction to Laplace's equation for steady-state thermal fields.

d'Alambert's and Fourier's (hence Terzaghi's) equations prove to have many points in common, in that both describe the physical processes by means of second partial derivatives, calculated against space; at the same time, they dramatically differ, as partial derivates referred to time differ by one order of magnitude. This difference in the structure translates d'Alambert's equation into solutions represented by elastic waves which propagate in space and time and Fourier's equation - reduced to Laplace's one for steady-state thermal fields - into harmonic functions, responsible for temperature diffusion in the medium.

When these elements are transferred to Terzaghi's equation, it occurs that this equation may be solved by combining d'Alambert's and Laplace's solutions, thus achieving a nonlinear equation which is capable of simulating the temporal evolution of excess pore pressures in the soil. Lastly, the research is completed with practical applications solved in parallel with Taylor's approximate solution, in order to take the advantages it provides, such as, among others, the independence from additional variables and the direct calculation of the evolution of excess pore pressures.

## 2  d'Alambert's Equation

d'Alambert's equation describes the oscillation of an elastic string, with a length $L$ and density $\rho$, subject to a force $F$:

$$\frac{\partial^2 \psi}{\partial x^2} = \frac{1}{v^2} \cdot \frac{\partial^2 \psi}{\partial t^2} \tag{1}$$

as a function of the propagation velocity of perturbation:

$$v = \sqrt{\frac{F}{\rho}} \tag{2}$$

The general solution:

$$\psi(x,t) = A \cdot \cos(k \cdot x + \delta) \cdot \sin(\omega \cdot t + \varphi) \tag{3}$$

describes the oscillation $\psi$ in the space 1D and in time through the wave number $k$, the constant phase $\phi$, the motion phase $\delta$, and the angular frequency $\omega$:

$$\omega = 2\pi \cdot f = \frac{2\pi}{T} \tag{4}$$

In short d'Alambert's equation is a set of solutions represented by elastic waves which propagate in space and time ($T \equiv t$), with constant phase velocity, carrying energy and motion quantity.

## 3   Fourier's and Laplace's Equations

In the case of 2D heat propagation, the issue is traced to Fourier's equation:

$$\frac{\partial^2 T_e}{\partial x^2} + \frac{\partial^2 T_e}{\partial z^2} = \frac{1}{\alpha} \cdot \frac{\partial T_e}{\partial t} \tag{5}$$

through the temperature ($T_e$) and the thermal diffusivity ratio $\alpha$.
By comparing eq. (5) with eq. (1) it is found out that both create a relation between the physical processes and space by means of second derivatives; at the same time, they dramatically differ by one order of magnitude, leading to solutions which are no longer comparable, although they offer important stimulus for thinking. Namely, for steady-state thermal conditions, Fourier's equation is reduced to Laplace's one:

$$\frac{\partial^2 T_e}{\partial x^2} + \frac{\partial^2 T_e}{\partial z^2} = 0 \tag{6}$$

describing the thermal field independently of the physical-thermal properties of the medium and containing closed form solutions:

$$T_e(x,z) = e^x \cdot \sin(z) \tag{7}$$

known as harmonic functions. Finally it may be noted that eq. (7) consisting in the application of an exponential function, damping the thermal wave, to the periodic functions in d'Alambert's eq. (3)

## 4   Exact Solution to Terzaghi's Equation

Terzaghi's equation, analogous to Fourier's eq. (5) reduced to the 1D case:

$$c_v \cdot \frac{\partial^2 u}{\partial z^2} = \frac{\partial u}{\partial t} \tag{8}$$

describes the spatial-temporal variation of pore pressures ($u$) through the primary consolidation coefficient:

$$c_v = \frac{K_z}{\gamma_w \cdot m_v} \tag{9}$$

dependent upon the permeability $K_z$, upon the volume weight of water $\gamma_w$ and the compressibility ratio $m_v$.

The equation (9) may be rewritten in another, fully equivalent, form as a function of the pore pressure variation:

$$c_v \cdot \frac{\partial^2 \Delta u}{\partial z^2} = \frac{\partial \Delta u}{\partial t} \tag{10}$$

By using the elements introduced above, it occurs that the research of the solution to eq. (8) or (10) begins with the application of a static load, on the surface of a generic soil, which as a result generates excess pore pressures (*Δu*) whose evolution in time may be described with a periodic function such as:

$$\Delta u(t) = \Delta u \cdot \cos(\omega \cdot t) \tag{11}$$

For eq. (11), as against the time factor alone and describing "pore pressure propagation", to represent the solution to Terzaghi's equation, it must also be referred to space:

$$\Delta u(z,t) = \Delta u \cdot \cos(\omega \cdot t - k_u \cdot z) \tag{12}$$

with the combination of the periodic terms present in eq. (3). Finally, taking advantage of the knowledge derived from eq. (7), it is found out that the solution to Terzaghi's equation may be achieved by introducing a damping function, dependent upon a variable $k_u$ and the depth $z$, to eq. (12):

$$\Delta u(z,t) = \Delta u \cdot e^{-k_u \cdot z} \cos(\omega \cdot t - k_u \cdot z) \tag{13}$$

By calculating first and second derivatives (please refer to Appendix for relevant details) and through the substitutions in eq. (10) the *consolidation variable $k_u$* may be formulated:

$$k_u = \sqrt{\frac{\omega}{2c_v}} = \sqrt{\frac{\pi}{c_v \cdot t}} = m^{-1} \tag{14}$$

this depends on angular pulse, eq. (4), and the primary consolidation coefficient. In other words, it must not be determined experimentally, as it is time dependent as against $c_v$.

Analysing in detail eq. (13) it turns out that it simulates the temporal variation of excess pore pressures in the soil, reducing its width as depth increases by means of an inverse hyperbolic function. In other words, pore pressure variation spreads in the ground, diminishing with distance, rather than propagating, as is the case, instead, in the vibratory motions from which part of the solution derives.

## 5  Practical Applications

The applications deriving from eq. (13) required an initial phase of in-depth study leading, at the moment, to two different uses, aimed at analysing the temporal evolution of the consolidation phenomenon.

Take for example a 20 m thick clay bank resting on gravels, that corresponds to a drainage path with a length H = 10 metres (Fig. 1a); on this bank a static load N = 100 kPa is applied, whereas the performance of oedometer tests showed that $c_v$ = 0.02 m²/day. Furthermore, it is assumed that load application times are negligible as against clay filtration times, thus leading to a system that initially is undrained and capable of developing a Δu = N = 100 kPa in relation to the incompressibility of both solid skeleton and pore fluid.

The first application of eq. (13) consists in the construction of "*n*" dissipation curves of Δu, in the example limited to $t_1$ = *2 years* ÷ $t_n$ = *20 years*, which are matched by the values of *ω* and $k_u$ in Fig. 1c. The values show a decrease of the latter when time increases, that is when subsidences increase and, as a result, when the void ratio decreases. Analysing in detail the graph of Fig. 1b it is found out that consolidation is completed via times close to the curve of 18 years (*U = 100%*) as it remains in the field of the Δu ≥ 0 and ceases at the depth of 10 metres, corresponding to the drainage length; simultaneously Taylor's solution (*U = 95% → $T_v$ = 1.129*) translates into $t_{95}$ = *15.5 years*.

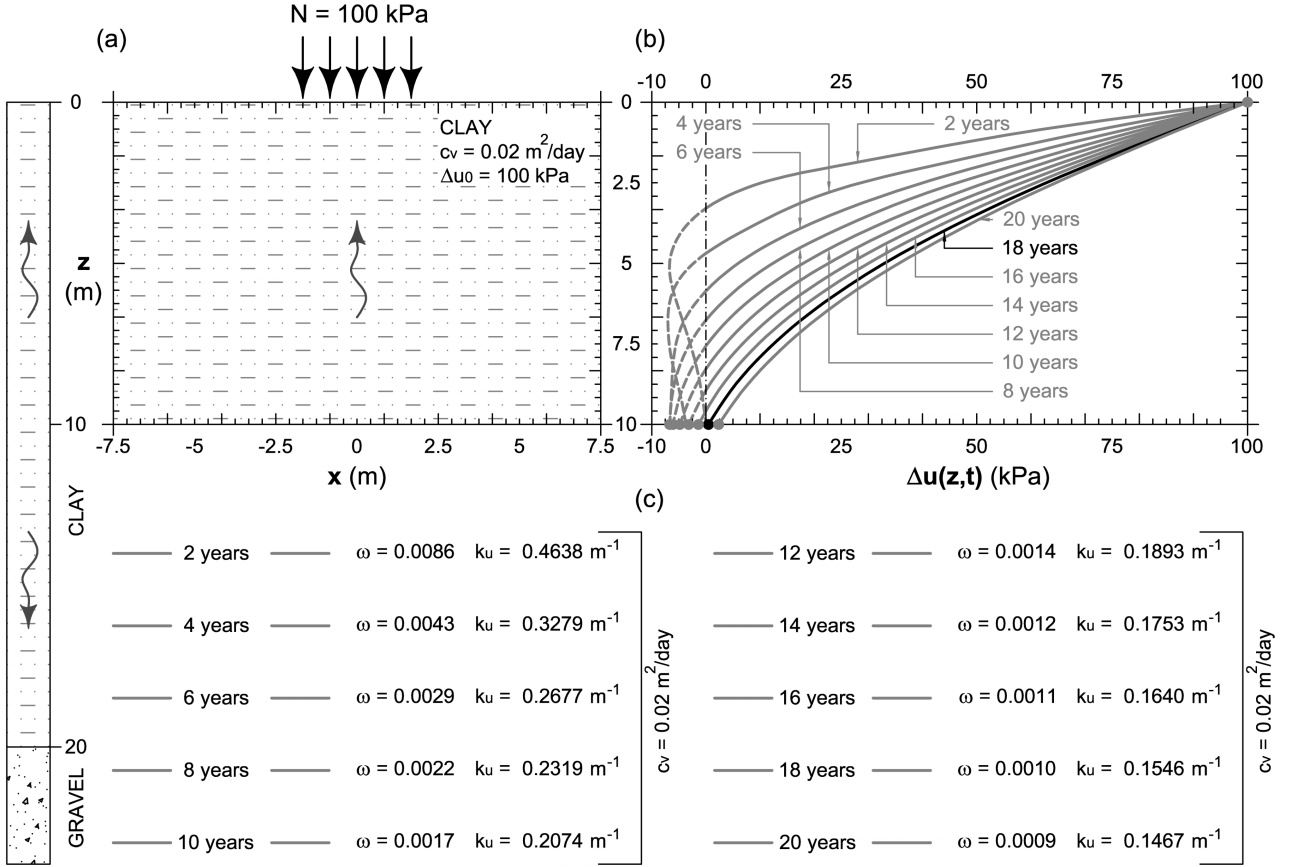

Fig. 1. Application example of the exact solution to Terzaghi's equation: (a) reference stratigraphy; (b) dissipation curves; (c) w, $k_u$.

The identification of the dissipation curve, corresponding to the completed consolidation for the condition:

$$\Delta u(z, t_{100}) = \Delta u \cdot e^{-k_u \cdot z} \cos(\omega \cdot t - k_u \cdot z) = 0 kPa \qquad (15)$$

leads to a second usage method for eq. (13) requiring the use of a math solver, such as Windows® Excel®, to look for the value of *t* that satisfies the eq. (15) at the analysis depth. By applying this method to the example in Fig. 1 you obtain a decay time $t_{100}$ = *17.5 years*, in line with the results obtained with the first method (t = 18 years) based on a few dissipation curves.

The procedure may then be completed through the construction of the curve *U ÷ t*, noting that each curve has a corresponding depth $z_i$ suppressing the relevant Δu (Fig. 2b), as is the case of the curve of completion of the phenomenon (t = 17.5 years) corresponding to $z_n$ = 10 meters; hence, considering it must be equal to:

$$U_{i(\%)} = \frac{z_i}{z_n} \cdot 100 \tag{16}$$

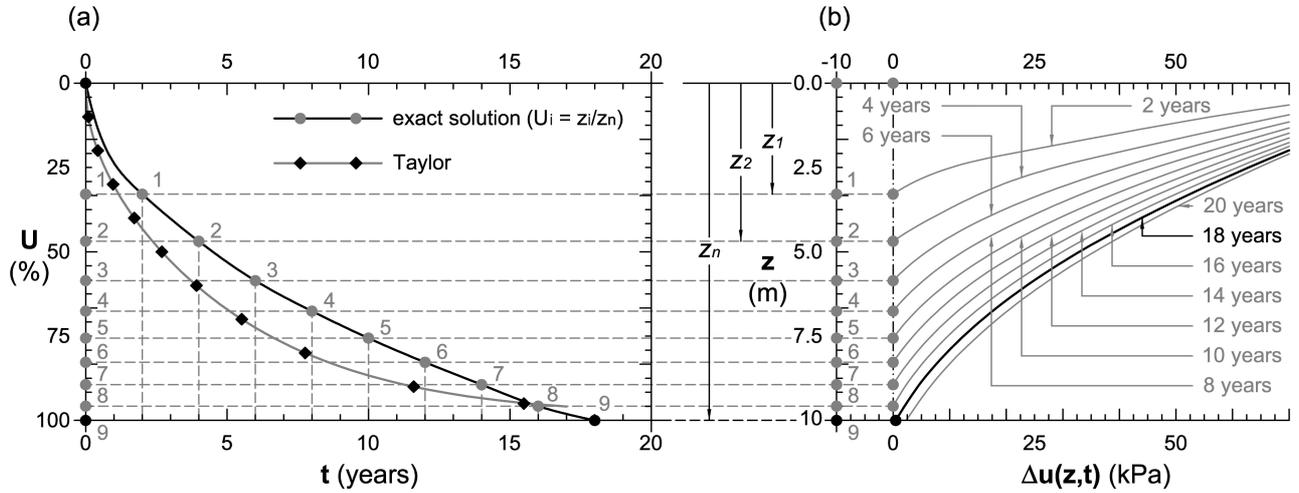

Fig. 2. (a) comparison between the curves U ÷ t related to the exact solution and to Taylor's approximate solution; (b) identification of suppression depths of Δu for every dissipation curve

U values corresponding to every dissipation curve may be identified leading to the graph in Fig. 2a. Not least, the variation of oedometric subsidences over time may derived from this curve.

# 6 Conclusions

The use of Terzaghi's equation is limited to Taylor's approximate solution based on the introduction of invariable adimensional parameters (Z, $T_v$), whatever the geological and tensional history of the soils.

This study illustrates the method followed in seeking the exact solution to Terzaghi's equation, whose starting point is given by the analysis of the time dependent behaviour of geomaterials, which may be described by means of hyperbolic functions, and of the logical structure of further differential equations, capable of describing some of the most important physical phenomena; thus it was found out that the solution sought ensues from the combination of partial solutions to d'Alambert's, Fourier 's, and Laplace's equations, combined with an inverse hyperbolic law, as a function of a new parameter (consolidation variable $k_u$) time-dependent upon the consolidation coefficient through the angular frequency. In other words, it turns out that the evolution of the consolidation phenomenon only depends on $c_v$ which, being derived experimentally from oedometer tests, summarises the previous tensional history of a given soil.

By applying the exact solution, performed concurrently with Taylor's approximate solution, a good correspondence with the decay times of the phenomenon may be detected and at the same time a slowing down along the entire consolidation curve showing a limit in the one-dimensional theory.

Please also consider that the exact solution may be reversed to determine $c_v$ starting from the experimental consolidation curves, as it is necessary to know the value of the Δu corresponding to $t_{100}$; hence, for $t_{100}$ = 510 sec, $H_0$ = 20 mm and $H_{100}$ = 16.57 mm, obtained from a real test, you obtain $c_v$ = 0.014 ÷ 0.024 ÷ 0.036 m²/day in the assumption that it works out to be $\Delta u_{100}$ = 0 ÷ 10 ÷ 20 kPa. When these results are applied to the example in Fig. 1 a consolidation decay time $t_{100}$ = 25 ÷ 14.6 ÷ 9.7 years is obtained, whereas from the same test, using the Casagrande method (1936), a $c_v$ = 0.032 m²/day was obtained, which, when included in Taylor's solution, provided a $t_{95}$ = 9.7 years.

In conclusion, it is essential to stress the need to explore the exact solution in the bi-three-dimensional field of consolidation, whereas it would be desirable to change oedometer cells in order to measure the performance of the Δu over time.

# Appendix

The validation of the exact solution requires the calculation of partial derivatives and their substitution in eq. (10). Deriving against time you have:

$$\frac{\partial \Delta u}{\partial t} = -\omega \cdot \Delta u \cdot e^{-k_u \cdot z} \cdot \sin(\omega \cdot t - k_u \cdot z) \tag{a}$$

whereas against space:

$$\frac{\partial \Delta u}{\partial z} = -k_u \cdot \Delta u \cdot e^{-k_u \cdot z} \cdot \cos(\omega \cdot t - k_u \cdot z) + k_u \cdot \Delta u \cdot e^{-k_u \cdot z} \cdot \sin(\omega \cdot t - k_u \cdot z) \tag{b}$$

$$\frac{\partial^2 \Delta u}{\partial z^2} = k_u^2 \cdot \Delta u \cdot e^{-k_u \cdot z} \cdot \cos(\omega \cdot t - k_u \cdot z) - k_u^2 \cdot \Delta u \cdot e^{-k_u \cdot z} \cdot \sin(\omega \cdot t - k_u \cdot z) -$$
$$+ k_u^2 \cdot \Delta u \cdot e^{-k_u \cdot z} \cdot \sin(\omega \cdot t - k_u \cdot z) - k_u^2 \cdot \Delta u \cdot e^{-k_u \cdot z} \cdot \cos(\omega \cdot t - k_u \cdot z) \tag{c}$$

Eq. (c) may be simplified:

$$\frac{\partial^2 \Delta u}{\partial z^2} = -2k_u^2 \cdot \Delta u \cdot e^{-k_u \cdot z} \cdot \sin(\omega \cdot t - k_u \cdot z) \tag{d}$$

and then introduced in eq. (10) together with eq. (a) obtaining at first:

$$c_v \cdot \left[ -2k_u^2 \cdot \Delta u \cdot e^{-k_u \cdot z} \cdot \sin(\omega \cdot t - k_u \cdot z) \right] = -\omega \cdot \Delta u \cdot e^{-k_u \cdot z} \cdot \sin(\omega \cdot t - k_u \cdot z) \tag{e}$$

and finally with a few mathematical manipulations:

$$k_u^2 = \frac{\omega}{2c_v} = \frac{\pi}{c_v \cdot t} \tag{f}$$

$$k_u = \sqrt{\frac{\omega}{2c_v}} = \sqrt{\frac{\pi}{c_v \cdot t}} \tag{g}$$